\documentclass[aps,prl,reprint,superscriptaddress]{revtex4-1}

\usepackage[dvipdfmx]{graphicx}
\usepackage{amsmath,booktabs}

\begin{document}


\title{Nonreciprocal Terahertz Second Harmonic Generation in Superconducting NbN under Supercurrent Injection}


\author{Sachiko Nakamura}
\email[]{snakamura@crc.u-tokyo.ac.jp}
\affiliation{Cryogenic Research Center, the University of Tokyo, Yayoi, Tokyo, 113-0032, Japan}
\author{Kota Katsumi}
\affiliation{Department of Physics, the University of Tokyo, Hongo, Tokyo, 113-0033, Japan}
\author{Hirotaka Terai}
\affiliation{National Institute of Information and Communications Technology, 588-2 Iwaoka, Nishi-ku, Kobe 651-2492, Japan}
\author{Ryo Shimano}
\email[]{shimano@phys.s.u-tokyo.ac.jp}
\affiliation{Cryogenic Research Center, the University of Tokyo, Yayoi, Tokyo, 113-0032, Japan}
\affiliation{Department of Physics, the University of Tokyo, Hongo, Tokyo, 113-0033, Japan}

\date{\today}

\begin{abstract}
Giant second-harmonic generation (SHG) in the terahertz (THz) frequency range is observed in a thin film of an s-wave superconductor NbN, where the time-reversal ($\mathcal{T}$-) and space-inversion ($\mathcal{P}$-) symmetries are simultaneously broken by supercurrent injection. 
We demonstrate that the phase of the second-harmonic (SH) signal flips when the direction of supercurrent is inverted, i.e., the signal is ascribed to the nonreciprocal response that occurs under broken $\mathcal{P}$- and $\mathcal{T}$-symmetries. 
The temperature dependence of the SH signal exhibits a sharp resonance, which is accounted for by the vortex motion driven by the THz electric field in an anharmonic pinning potential.  
The maximum conversion ratio $\eta_{\mathrm{SHG}}$ reaches $\approx10^{-2}$ in a thin film NbN with the thickness of 25\,nm after the field cooling with a very small magnetic field of $\approx1$\,Oe, for a relatively weak incident THz electric field of 2.8\,kV/cm at 0.48\,THz. 
\end{abstract}


\maketitle

Nonreciprocal electric transport phenomena, 
where electrical conductivity depends on the direction of the current ($I$), play crucial roles in modern electronics. 
Most typical example is a diode consisted of a p-n junctions, where the space-inversion ($\mathcal{P}$-) symmetry is \textit{macroscopically} broken.  
The nonreciprocal phenomena are found to emerge even in bulk materials where $\mathcal{P}$-symmetry is \textit{microscopically} broken~\cite{PhysRevLett.87.236602}, and the nonreciprocity can become prominent especially in the diagonal components of the conductivity $\sigma$ (i.e. $\sigma_{xx}(I)-\sigma_{xx}(0) \propto I$). 
The extension of the phenomena to higher frequency ranges such as microwave, terahertz (THz), and optical frequencies is highly demanded for high-speed optoelectronic devices~\cite{[][{, and references therein.}]Nat.Comm.9.3740,PhysRevApplied.10.047001}. 
For the linear and diagonal nonreciprocity, the time-reversal ($\mathcal{T}$-) symmetry should also be broken in addition to the $\mathcal{P}$-symmetry breaking according to the Onsager's reciprocal theorem~\cite{PhysRev.37.405}. 
Generally, the $\mathcal{T}$-symmetry breaking is induced by applying external magnetic fields or in the presence of magnetization~\cite{PhysRevLett.94.016601,10.1038/nphys3356,
PhysRevLett.117.127202,giantSHG2019nature,NPhys13.578,PhysRevB.92.184419,
NComm5.3757,doi:10.1063/1.1523895,SciAdv3.e1602390,NatComm8.14465,NatComm10.2734,Itahashieaay9120,PhysRevResearch.2.023127}. 
In superconductors, the nonreciprocal transport phenomena arising from various mechanisms have been investigated recently;
trigonal/hexagonal warping of band structures~\cite{SciAdv3.e1602390,PhysRevB.98.054510}, anisotropic friction forces on flowing vortices~\cite{PhysRevB.98.054510}, and parity mixing ~\cite{PhysRevLett.121.026601}
A recent theoretical study has even extended to nonlinear optical responses in noncentrosymmetric superconductors~\cite{PhysRevB.100.220501}.
Conversely, these nonreciprocal responses can be used as a probe to detect the small $\mathcal{P}$-symmetry breaking in noncentrosymmetric superconductors. 

\begin{figure}[b]
\includegraphics[width=0.4\textwidth]{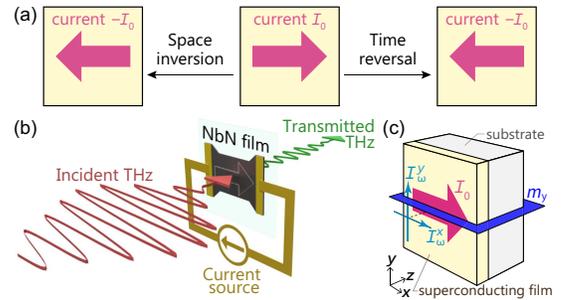}%
\caption{(Color online) (a) The dc supercurrent $I_0$ indicated by the arrow changes the direction under each time-reversal and space-inversion operations. (b) Schematic view of the terahertz transmittance experiments under the dc supercurrent. (c) Configuration of the sample and currents. \label{fig1}}
\end{figure}

In this Letter, we study the nonreciprocal electromagnetic responses in s-wave superconductors in the quasi-optical THz frequency range under dc supercurrent bias. 
Notably, the non-dissipative supercurrent breaks both of the $\mathcal{T}$- and $\mathcal{P}$-symmetries as depicted in Fig.~\ref{fig1}(a), and thus can be viewed as a sort of symmetry-breaking field in microscopic considerations. 
It should be noted here that the presented scheme makes a stark contrast to normal current injection where the dissipation plays a role of macroscopic irreversibility. 
We studied a thin film of s-wave superconductor NbN as a testing ground which can carry very large dc currents ($>1$\,MA/cm$^2$) without accompanying dc electric field due to zero resistance.
In addition, the induced magnetic field is expelled from the sample as it is smaller than the lower critical field, and other properties such as optical conductivities and superfluid densities are almost unaffected as long as the current is kept under the critical value~\cite{PhysRevLett.122.257001}. 
In the presence of a dc electric supercurrent $I_0$, we observed nonreciprocal second harmonic generation (SHG) in response to the oscillating current $I_\omega$ at THz frequencies. 
The SHG signal shows the phase inversion when the direction of dc current is reversed, which is the hallmark of the linear nonreciprocal response, e.g., $\sigma_{xx}(I_\omega,I_0)-\sigma_{xx}(0,0) \propto I_\omega I_0$, arising from the current-induced $\mathcal{T}$-symmetry breaking. 

The SHG was measured using a THz time-domain spectroscopy (THz-TDS) technique in a transmission geometry as schematically illustrated in Fig.~\ref{fig1}(b), to realize the phase-resolved detection. 
The intense monocycle THz pulse was generated by the tilted-pulse front method with a LiNbO$_3$~\cite{Hebling:02,Watanabe:11} and spectrally narrowed by using band-pass filters with different center frequencies. 
Typical peak values of the electric field were 2.8, 2.7, 4.0, and 4.6\,kV/cm for 0.48, 0.6, 0.8, and 1\,THz sources, respectively. 
A regenerative amplified Ti:sapphire laser system with 800\,nm center wavelength, 100\,fs pulse duration, 1\,kHz repetition rate, and pulse energy of 1 or 4\,mJ was used as a light source. 
The THz intensity and polarization angle were controlled with wire grid polarizers. 
The transmitted THz pulse after the sample was detected by electro-optic (EO) sampling using a ZnTe crystal. 
The superconducting film was an epitaxial NbN films of $25$\,nm in thickness grown on a 400-$\mu$m-thick MgO (100) substrate. 
The dc current was injected into the 4-mm-square area through the Au/Ti electrodes deposited on the surface as shown in Fig.~\ref{fig1}(b), and THz pulse was focused on its center where the dc current flows almost uniformly~\cite{PhysRevB.48.12893}.  
The critical temperature $T_\mathrm{c}$ is $14.5\pm0.2$\,K and the critical current $I_\mathrm{c}$ is 3.3\,A (3.3\,MA/cm$^2$) at $T=5$\,K. 
The critical current is defined as the current where the quench occurs by increasing the current by 0.01\,A with 5\,sec intervals. 

\begin{figure}[bt]
\includegraphics[width=0.48\textwidth]{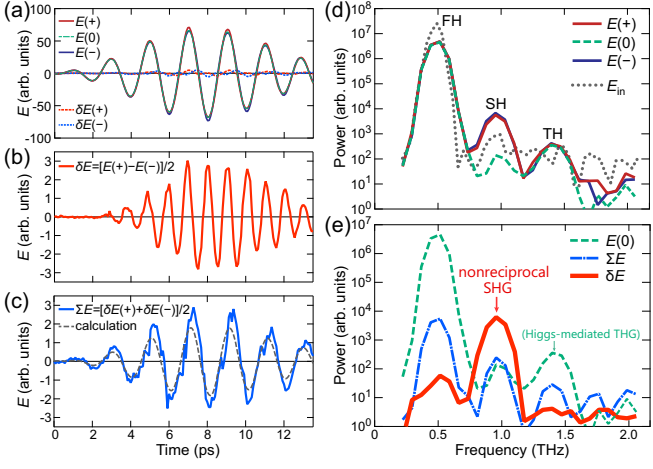}%
\caption{(Color online) THz transmission measurements on a NbN film at $T=11.6$\,K and $\omega/2\pi=0.48$\,THz (peak value: 5.4\,kV/cm). (a)--(c) Real-time waveform of the transmitted THz electric fields $E(\pm)$ and $E(0)$ measured with $I_0=\pm 2.1$ and $0$\,A, respectively, and current-induced changes $\delta E(\pm)\equiv E(\pm)-E(0)$, $\delta E\equiv \left[E(+)-E(-)\right]/2$, and $\Sigma E\equiv \left[\delta E(+)+\delta E(-)\right]/2$. The dashed line in (c) shows the expected change due to the current-induced optical conductivity change~\cite{PhysRevLett.122.257001}. 
(d),(e) Power spectra of the transmitted THz electric fields and incident THz electric field $E_\mathrm{in}$. The FH, SH, and TH in (d) stand for the fundamental, second, and third harmonics. Power spectra of the current induced changes are shown in (e).\label{fig2}}
\end{figure}

Figure~\ref{fig2}(a) shows waveforms of the transmitted THz pulse through the NbN film with a dc current in one direction $E(+)$, in the opposite direction $E(-)$, and without current $E(0)$ at $T=11.6$\,K. 
The polarization of the incident THz electric field is parallel to the dc current. 
The center frequency of the incident THz pulse $\omega/2\pi$ is 0.48\,THz, and the peak value of the incident THz field is 5.4\,kV/cm. 
Figure~\ref{fig2}(d) shows the power spectra of the waveforms. 
Without the current injection, the power spectrum of the transmitted THz pulse [$E(0)$] has a peak at the fundamental harmonic (FH) frequency, and the small peak observed at the second harmonic (SH) frequency $2\omega$ is attributed to the transmission of the leakage of the band-pass filter which is inserted before the film to narrow the incident FH. 
A peak at the third harmonic (TH) frequency $3\omega$ originates from the Higgs mode~\cite{Matsunaga1145}. 
By injecting a dc current of 2.1\,A, a clear peak appears at $2\omega$, whose intensity is $10^{-3}$ of the FH intensity. 
Compared to the power spectrum of $E(0)$, the signal to noise ratio for this peak is about 50. 
The SH intensity is larger than the tail component of the incident field [$E_\mathrm{in}$ in Fig.~\ref{fig2}(d)] at the frequency $2\omega$ by a factor of eight, which confirms that the peak is attributed to SHG and not due to the current-induced transmittance change. 
The SH intensity is proportional to square of the FH intensity (shown in Fig.~S2(a) in Supplemental Material), exhibiting the fundamental characteristics of second-order nonlinear phenomena. 
The SHG conversion ratio (up to $10^{-2}$~\footnote{Fig.~S1(a) in Supplemental Material.}) is extraordinarily large for a 25-nm-thick film. 
The observed conversion efficiency  corresponds to the effective nonlinear susceptibility, ${\chi^{(2)}_\mathrm{eff}}\approx 2.2\times 10^{8}$\,pm/V, if we dare to plug the value into the conventional expression for the second order nonlinear susceptibility of nonlinear optical crystals, $\chi^{(2)}_\mathrm{eff}=\sqrt{\dfrac{\mathrm{SH}}{\mathrm{FH}}\dfrac{n_\omega n_{2\omega} {\lambda_{2\omega}}^2}{d^2 \pi^2 |E_\omega|^2}}$~\cite{doi:10.1002/adom.201500723} where the refractive indices $n_\omega$ and $n_{2\omega}$ are assumed to be $\approx 1$, $d$ is the crystal thickness, and $\lambda_{2\omega}$ is the SH wavelength.  
This value is about $10^4$ times larger than that of LiTaO$_3$, a material known to possess a large $\chi^{(2)}$ in the THz frequency range~\cite{PhysRevB.33.6954}. 
Figure~\ref{fig2}(b) clearly manifests that the SH ($2\omega$) oscillation arises from the nonreciprocal part of the current-induced signal, $\delta E \equiv \left[E(+)-E(-)\right]/2$, whereas the reciprocal part $\Sigma E \equiv \left[\delta E(+)+\delta E(-)\right]/2$ oscillates in $\omega$ as shown in Fig.~\ref{fig2}(c). 
The observed reciprocal part $\Sigma E$ shows a good accordance with the calculated one and thus is attributed to the current-induced conductivity change~\cite{PhysRevLett.122.257001} as plotted by a dashed line in Fig.~\ref{fig2}(c). 
The small peak at $2\omega$ appeared in the power spectrum of $\Sigma E$ originates from the leakage of the filter. 

\begin{figure}[bt]
\includegraphics[width=0.45\textwidth]{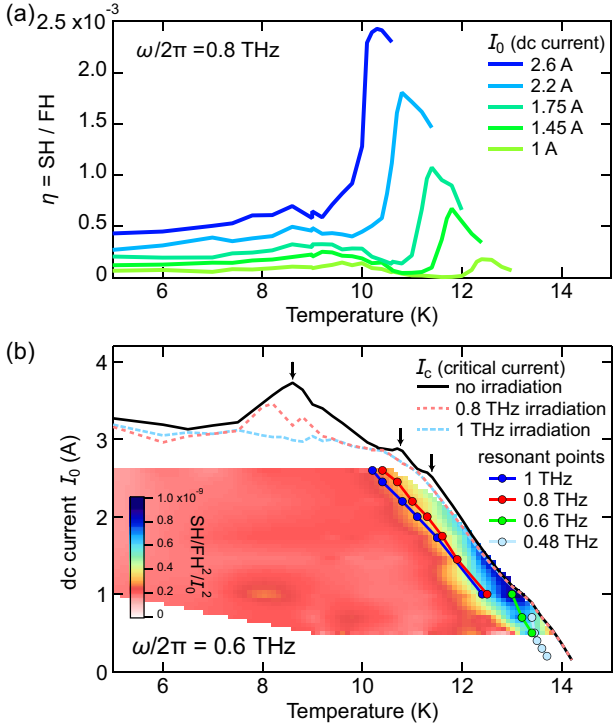}%
\caption{(Color online) (a) Temperature ($T$) dependence of SH conversion factor $\eta$. (b) $T$-$I_0$ phase diagram. The color map shows SHG enhancement factor (defined in the text) as a function of the $T$ and $I_0$.\label{fig3}}
\end{figure}

The temperature dependence of the SH conversion efficiency $\eta$ (intensity ratio of SH and FH) is represented in Fig.~\ref{fig3}(a). 
It is nearly constant in the low temperature range, and exhibits a prominent peak at high temperature which shifts to lower temperature side with increasing the dc current $I_0$. 
Regardless of the temperature, $\eta$ shows a quadratic dependence to the dc current $I_0$~\footnote{Fig.~S2(b) in Supplemental Material.}. 
A similar peak is found for other incident fundamental frequencies $\omega$, while the peak height strongly depends on the incident frequency (see Fig.~S1 in Supplementary Materials). 
When the probe THz polarization is perpendicular ($\perp$) to the dc current, the SH intensity is smaller than that with parallel configuration ($\parallel$) by more than two-orders of magnitude~\footnote{Fig.~S2(c) in Supplemental Material.}. 
This result is reasonable since there exists a mirror symmetry $m_y$ [Fig.~\ref{fig1}(c)] where $I_0$ goes along the x-axis, with which the linear and diagonal nonreciprocity for the oscillating current along the y-axis ($I_{\omega}^{y}$) is forbidden. 
Figure~\ref{fig3}(b) represents the color map of SHG enhancement factor as defined by SH/FH$^2/{I_0}^2$ for $\omega/2\pi=$0.6\,THz as a function of temperature $T$ and dc current $I_0$. 
The boundary of the superconducting phase and the normal metal phase is indicated by the black solid curve which identifies the critical current $I_\mathrm{c}$. 
Clearly, the SHG is enhanced in the vicinity of the $I_\mathrm{c}$ boundary.
The resonant peak positions observed in Fig.~\ref{fig3}(a) for $\omega/2\pi=0.8$\,THz source and for other frequencies are plotted by closed circles in the phase diagram of Fig.~\ref{fig3}(b). 
The resonant points appear slightly below the phase boundary, and are located inward for higher $\omega$. 
The observed current-induced SHG accompanied by a characteristic resonance behavior can be phenomenologically described by an anharmonic oscillator model~\cite{anharmonicmodel1965} where the $\mathcal{P}$-symmetry breaking is induced by the dc supercurrent as discussed in detail in the following. 

\begin{figure}[bt]
\includegraphics[width=0.48\textwidth]{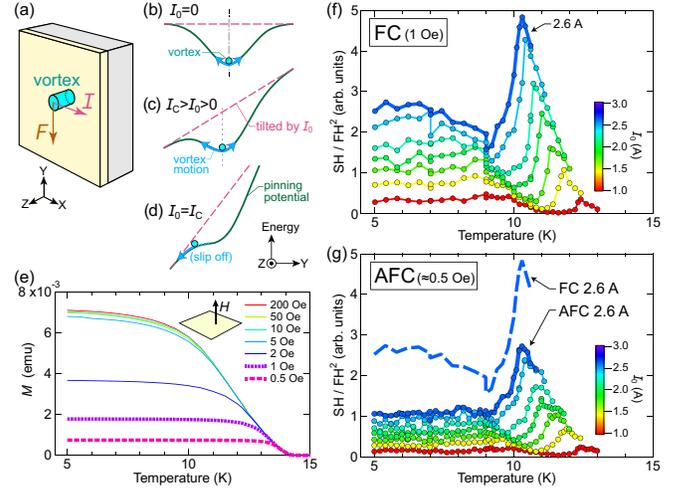}%
\caption{(Color online) (a) Schematic view of a vortex normal to the surface subjected to a force $F$ along Y-axis due to a current $I$ along X-axis. Corresponding pinning potentials without dc current ($I_0=0$), with current ($I_0>0$), and with the critical current ($I_0=I_\mathrm{c}$) are shown in (b), (c), and (d), respectively. (e) Trapped magnetization $M$ measured with increasing temperature at $H=0$\,Oe after cooling from $T=20$\,K to 5\,K across $T_\mathrm{c}=14.5$\,K in the indicated preparation field ($H_\mathrm{pre}=$0.5, 1, ..., 200\,Oe) applied normal to the film.  (f, g) SHG intensity scaled by FH intensity for $\omega/2\pi$=1\,THz (f) after cooling with permanent magnets (FC, 1\,Oe) and (g) in the ambient field (AFC). \label{fig4}}
\end{figure}

As a microscopic origin of the anharmonic oscillator, here we consider the motion of pinned vortices induced by the remnant magnetic field.
In type-II superconductors, the critical current $I_\mathrm{c}$ corresponds to the depinning current where the pinning force balances with the Lorentz force $F$~\cite{PhysRevLett.16.734} due to the dc current $I_0$. 
Figure~\ref{fig4}(a) shows a typical configuration of a vortex, current, and the force in the film. 
Without the dc currents, the vortex is trapped in a symmetric pinning potential $V_\mathrm{pin}$ as schematically shown in Fig.~\ref{fig4}(b). 
When the dc current ($I_0$) is applied, the pinning potential is tilted and becomes asymmetric as shown in Fig.~\ref{fig4}(c), and the equilibrium point shifts to the left. 
When the current reaches $I_\mathrm{c}$, the vortex slips off the pinning potential as shown in Fig.~\ref{fig4}(d). 
Then, the incident THz field induces an oscillating current $I_\omega$ along the X-axis which causes the oscillation of the vortex along the Y-axis. 
The equation of motion for a vortex along the Y-axis is given by~\cite{jisokuSM} 
\begin{equation}\label{eq:EOM1}
m\frac{\mathrm{d} v}{\mathrm{d} t}
=-\frac{\partial }{\partial y}\left(V_\mathrm{pin}+J_0 d_\mathrm{f} \Phi_0 y\right) - \eta d_\mathrm{f} v - d_\mathrm{f} \Phi_0 J_\omega 
\end{equation}
where $m$, $v$, and $y$ are the mass, velocity, and position of the vortex, respectively, $\Phi_0$ is the flux quantum, $d_\mathrm{f}$ is the thickness of the film, and $J_i$ ($i=0, \omega$) is the current density corresponding to $I_i$. 
The first term of eq.~(\ref{eq:EOM1}) is the restoring force due to the $V_\mathrm{pin}$ tilted by $J_0$, the second term is the viscous drag, and the last term is the driving force oscillating at $\omega$. 
If there are higher harmonic components in $V_\mathrm{pin}$ (i.e., $y^4, y^6, ...$), the Taylor series of the potential around the new equilibrium point has odd power terms which gives rise to even harmonics oscillation along the Y-axis. 
Because the moving vortex induces electric field perpendicular to the vortex motion $\vec{E}_\mathrm{ind}=\Phi_0 \vec{z} \times \vec{v}$~\cite{JOSEPHSON1965242,doi:10.1143/JPSJ.57.3941}, the even harmonics of electromagnetic field is expected to be emitted along the X-axis, which is parallel to the incident THz. 
According to this mechanism, the SH intensity should be proportional to $\mathrm{FH}^2$, $I_0^2$, and square of the vortex density. 
The first two features are consistent with our observations. 
To examine this picture further, we tried to change the vortex density. 
 
In our configuration, the vortex density is determined by the applied magnetic field normal to the film when the sample is cooled across $T_\mathrm{c}$. 
The magnetic field induced by the dc current, which is parallel to the film, is expelled from the film~\footnote{The current-induced field ($<5$\,Oe) is much smaller than the in-plane penetration field ($H^{\parallel}_\mathrm{p}\approx 100$\,Oe) measured for another sample fabricated in very similar conditions, which is found in Supplemental Material, which includes Refs.~\cite{doi:10.1063/1.1661678,pinningMatsushita}.}. 
Figure~\ref{fig4}(e) shows the temperature dependence of the magnetization $M$ of the NbN film which was first cooled under the presence of indicated magnetic fields $H_\mathrm{pre}$, and then warmed up without the field. 
The magnetization $M$ corresponds to the total magnetic flux of pinned vortices, i.e., the vortex density, which increases by increasing the $H_\mathrm{pre}$ and saturates at $H_\mathrm{pre}\approx5$\,Oe for $T=5$\,K and $H_\mathrm{pre}\approx 1$\,Oe for $T=13$\,K. 
We tried field cooling (FC) with applying an external magnetic field of $H_\mathrm{pre}\approx$1\,Oe using a pair of neodymium magnets placed outside the cryostat, and compare with ambient field cooling (AFC) where $H_\mathrm{pre}\approx 0.5$\,Oe. 
As the external magnetic field suppresses the $I_\mathrm{c}$, we removed the magnets before the measurements so that the $I_\mathrm{c}$ comes back to the original value. 
The SH intensity is clearly enhanced by FC compared to that for AFC without changing the resonance temperature, as shown in Figs.~\ref{fig4}(f) and (g). 
Because the FC under the very weak field changes only the density of the frozen vortices, this observation indicates that the nonreciprocal SH originates from the pinned vortices in the sample. 
According to the mechanism proposed above, the observed decrease of the resonant frequency with increasing the $T$ or $I_0$ is accounted for by the shallowing of the effective pinning potential. 

In the phase diagram shown in Fig.~\ref{fig3}(b), the critical current $I_\mathrm{c}$ gradually decreases with increasing the temperature. 
Notably, one can see peaks at 8.5, 10.8, and 11.2\,K on the $I_\mathrm{c}$ curve marked by the arrows, which suggests that the vortices slowly creep to optimize the vortex lattice and deepen the effective pinning potential at substantially high temperature~\cite{PhysRevLett.9.306,PhysRevLett.9.309}. 
Under the THz irradiation, as plotted by broken lines in the same figure, the peaks are clearly suppressed, which suggests that the vortices are certainly oscillated by the THz irradiation and slip out of the pinning potential before optimizing their configuration. 
Having seen that the experimental results are consistently described by the vortex motion, based on eq.~(\ref{eq:EOM1}), we evaluate the parameters, i.e., vortex mass ($m\approx m_e$ where $m_e$ is the electron mass), viscosity [$\eta\approx 10^{-10}$\,kg /(sec m)], and eigenfrequency of the vortex motion (in the THz frequency range), from the current and temperature dependence of the resonant frequency~\cite{jisokuSM}. 
The mass per unit length $m/d$ is estimated as $\approx 10^7 m_e$/meter where $d$ is the film thickness, 25\,nm. 
This value shows a reasonable agreement with that calculated for the vortex core in dirty limit superconductors ($\approx 10^7 m_e$/meter)~\cite{PhysRevLett.14.226,PhysRevB.57.575}, whereas previous experiments have reported much heavier masses, $\approx 10^{12}$ and $10^{10} m_e$/meter~\cite{doi:10.1063/1.2747080,PhysRevB.85.060504}. 
The difference suggests that the high frequency THz response in the present experiment is sensitive to the bare mass of vortex core and less affected by dissipative dynamics~\cite{doi:10.1063/1.2747080,PhysRevB.85.060504,RevModPhys.66.1125}.
The observed SH electric field is also in reasonable accordance with the calculated value of 140\,V/m, which is estimated, e.g. for $T=$10.4\,K, $\omega/2\pi=0.8$\,THz, and $I_0=2.6$\,A~\cite{jisokuSM}. 
Because the SH photons have higher energy than the superconducting gap, they can excite quasi-particles in the superconducting film. However, the observed SH field is much smaller than the threshold field for identifiable conductivity change, $\approx 10$\,kV/cm~\cite{PhysRevLett.109.187002}, and the quasi-particle density expected for the observed SH field is negligibly small ($<10^{-4}$ of the superfluid density). 

Finally, we address the contribution of Higgs mode in the SH signal. 
In the presence of dc current, the Higgs mode is directly excited by the fundamental wave ($\omega$)~\cite{PhysRevLett.118.047001, PhysRevLett.122.257001, PhysRevB.101.220507} which in turn should induce the SHG. 
The intensity of this Higgs-mediated SHG is expected to be comparable with that of the Higgs-mediated THG [shown in Fig.~\ref{fig2}(e)] because $I_0$ is as large as $I_\omega$\footnote{$I_\omega$ is calculated from the optical conductivity of the film and the internal electric field $E_0$ considering the multiple reflections.}.
In the present case, the Higgs-mediated SHG is overwhelmed by much stronger SHG induced by the vortex motion~\footnote{The $\eta$ for $\omega=0.48$\,THz shown in Fig.~S1(a) of Supplemental Material shows a second peak for $I_0=0.5$ and 0.7\,A at $T=13.2$\,K where the resonance condition $\omega=2\Delta$ is fulfilled, which might be related to the Higgs mode.}. 
Recently, SHG has been observed in a clean limit superconductor where the $\mathcal{P}$-symmetry breaking is caused by the effect of THz field-induced persistent photocurrent~\cite{NatPhys13.707,PhysRevLett.124.207003}. 
The method is advantageous for extracting ultrafast $\mathcal{P}$-symmetry breaking phenomena which can follow the time scale of THz wave cycle. 
In contrast, the static bias of dc supercurrent enables the investigation of nonreciprocal responses in a $\mathcal{P}$- and $\mathcal{T}$-symmetry broken system under a nonequilibrium steady state that accompanies slow dynamics, e.g., vortex reconfiguration, lattice distortion, or thermal relaxation. 

In summary, we have demonstrated the giant THz SHG in a type-II superconducting film of NbN under dc supercurrent injection. 
Based on the orientation dependence on the dc current and THz electric field polarization, we elucidate that the dc current breaks the time-reversal and space-inversion symmetries and gives rise to a colossal nonreciprocal SHG. 
This phenomenon would pave a new pathway for nonreciprocal THz electronics with superconductors. 
The observed maximum SH conversion efficiency reaches 1\% for a few kV/cm of THz electric field input even for the film sample of 25\,nm in thickness.  
The SH is enhanced by the field cooling with the resonant peak always appearing slightly below the critical current line in the temperature-current phase diagram. 
A microscopic model taking into account the vortex dynamics is shown to describe well the observed experimental behaviors, indicating that the microscopic origin of the SHG is attributed to the motion of frozen vortices in anharmonic pinning potential that is tilted by the dc current. 
While the broken inversion symmetry is induced by the supercurrent injection in the present work,  the demonstrated scheme of nonreciprocal THz-SHG response would be also promising for the study of superconductors with nonsymmetric order parameters. 

\begin{acknowledgments}
This work was supported in part by JSPS KAKENHI (Grants No.~15H02102, No.~18K13496, and No.~20K14408), and by JST CREST Grant Number JPMJCR19T3, Japan. 
\end{acknowledgments}

%

\end{document}